\documentclass[twocolumn,showpacs,preprintnumbers,amsmath,amssymb]{revtex4}

\usepackage{graphicx}
\usepackage{dcolumn}
\usepackage{bm}

\begin{document}

\preprint{APS/123-QED}


\title{Thickness-tuned Superconductor-to-Insulator
Transitions under magnetic field in a-NbSi}


\author{C.A. Marrache-Kikuchi}
\email{Claire.Marrache@csnsm.in2p3.fr}
\affiliation{%
CSNSM (CNRS-UMR8609), Universit\'{e} Paris Sud, Bat. 108, 91405 Orsay Campus, France
}%

\author{H. Aubin}
\author{A. Pourret}
\author{K. Behnia}
\author{J. Lesueur}%
\affiliation{Laboratoire Photons et Mati\`{e}re (CNRS), ESPCI,
10 rue Vauquelin, 75231 Paris, France
}%

\author{L. Berg\'{e}}
\author{L. Dumoulin}
\affiliation{%
CSNSM (CNRS-UMR8609), Universit\'{e} Paris Sud, Bat. 108, 91405 Orsay Campus, France}%

\date{\today}


\begin{abstract}
We have studied the thickness-induced superconductor-to-insulator transition
in the presence of a magnetic field for a-NbSi thin films. Analyzing the
critical behavior of this system within the "dirty boson model", we have
found a critical exponents product of $\nu_d z \sim 0.4$. The corresponding
phase diagram in the ($H$,$d$) plane is inferred. This small exponent product
as well as the non-universal value of the critical resistance found at the transition call for
further investigations in order to thoroughly understand these transitions.
\end{abstract}








\maketitle

\section{Introduction}

    Low temperature transport in disordered conducting materials imply
quantum interferences, Coulomb repulsion, and superconducting
fluctuations. Since 2D is the lower critical dimension for the
existence of both the superconducting and the metallic states,
transport properties of such disordered thin films have attracted
continuous attention since the 1960s in order to understand what ground states are allowed in those systems and study
the nature of the quantum phase transitions between the different phases \cite{Sondhi1997,Fisher1990,Fisher1990bis}.

    Quantum Phase Transitions (QPT) occur when a parameter in the
Hamiltonian is varied, resulting in a change of the system's ground
state. These transitions therefore take place at zero temperature
and are driven by quantum fluctuations, contrary to classical phase
transitions which are controlled by thermal fluctuations. Near a
QPT, the quantum fluctuations have a characteristic lengthscale -
the correlation length $\xi$ - diverging as $\xi \propto
\delta^{-\nu}$ where $\nu$ is the correlation length critical
exponent, $\delta_K = \frac{|K-K_{c}|}{K_c}$ the distance of the
considered system to the $K$-driven transition, and $K$ an experimentally tunable
parameter which critical value is $K_c$. The fluctuations are also
characterized by a vanishing frequency $\Omega \propto \xi^{-z}$
where $z$ is the dynamical critical exponent. The two critical
exponents $\nu$ and $z$ define the universality class to which the
transition belongs.

    In the case of Superconductor to Insulator Transitions (SITs) in disordered thin films, the tunable parameter
in the Hamiltonian can be the disorder or the magnetic field
$H$. The most popular theoretical model to explain these SITs is the
"dirty boson model" developed by M.P.A. Fisher \cite{Fisher1990}. In
this model, the coherence of the superconducting state is destroyed
by quantum fluctuations of the order parameter's phase
and the system amounts to interacting bosons
in the presence of disorder. The superconducting and insulating
phases are then dual to one another : the superconducting phase
consists of localized vortices and condensed Cooper pairs, whereas
the insulating phase is characterized by condensed vortices and
localized Cooper pairs. Both disorder and magnetic field-driven
transitions have similar description within this frame : in the
quantum regime, for DC measurements, the sheet resistance obeys a
scaling law that is solely dependent on the variable $\delta \ast
T^{-\frac{1}{\nu z}}$ \cite{Sondhi1997,Fisher1990}:

        \begin{equation}
        \label{eq:loi_échelle_B_d}
            R(\delta,T) = R_{c}f(\alpha \delta T^{-\frac{1}{\nu z}})
        \end{equation}

\noindent where $R_{c}$ is the critical sheet resistance and $f$ an
universal scaling function having an unique constraint : $f(0) = 1$.
$\alpha$ is a non universal constant \cite{Cha1991}.
$z = 1$ is expected due to the long-range Coulomb interactions
and the "dirty boson model" predicts $\nu >
\frac{2}{d} = 1$ as well as an universal value of the system's sheet
resistance at the transition $R_c=R_Q=\frac{h}{4e^{2}} =
6500$\, $\Omega$ \cite{Fisher1990bis}. Despite obeying to the same scaling
laws (equation (\ref{eq:loi_échelle_B_d})), the field-induced transition and the disorder-induced
transitions have different physical grounds : in the magnetic
field-induced SIT, the vortex density increases with the magnetic
field, until they delocalize and Bose condense ; in the
disorder-induced SIT at zero field, the Bose condensation is
undergone by the vortex/antivortex pairs. These two SITs hence have
no reason to have the same critical exponents \cite{Fisher1990bis}.

    Experimentally, number of disordered superconducting films
experience a SIT when submitted to a perpendicular magnetic field.
However, they do not all behave in the same way. Following
Gantmakher's comment \cite{Gantmakher2004}, one can separate them
into two different categories. Some compounds exhibit an insulating phase which low temperature resistance
is only 10\% above their high temperature resistance. This behavior
resembles more the one of a conductor in the
presence of weak localization than the one of an actual insulator \cite{Steiner2005}. This is the case of Mo$_x$Ge$_{1-x}$
\cite{Yazdani1995}, Mo$_x$Si$_{1-x}$ \cite{Okuma1998}, Be
\cite{Bielejec2002}, a-Bi \cite{Markovic1999}, or
Nd$_{2-x}$Ce$_x$CuO$_{4+y}$ \cite{Gantmakher2004}. Other systems, such as
amorphous indium oxyde \cite{Gantmakher2000ter} or TiN \cite{Baturina2005}, have, in the same conditions, a much more
important increase in resistance - up to a factor 10. Their
resistance then have an exponential increase with the temperature
\cite{Hebard1990,Gantmakher2000ter}. The renormalization analysis of these field-induced
SIT gives $0.75 \leq \nu_H z \leq 1.35$,
independently of the above-mentioned categories.

    The experimental realizations of the thickness-induced SIT,
where tuning the system's thickness is taken to be a mean of varying
its disorder, are far more rare because of the experimental
difficulty of synthesizing microscopically identical films which only
differ by their thicknesses. In the case of this transition, the
distinction previously made does no longer exist : all studied
compounds show a drastic increase in resistance of many orders in
magnitude when their thickness is lowered \cite{Markovic1999}.
However, one can make another distinction. Some systems, such as a-Bi
\cite{Markovic1999}, are very sensitive to any thickness variation : a fraction of
angstr\"{o}m difference engenders resistance increases of several orders of
magnitude at low temperature. This behavior is comparable to the one
observed in granular systems \cite{Jaeger1989}. On the other hand, systems such as
MoC present a more progressive thickness-dependence \cite{Lee1990}. Values of
the critical exponents have only been reported for a-Bi \cite{Markovic1999} :
$\nu_d z \sim 1.3$.

    Whichever the parameter tuned to induce the SIT, and contrary
to the predictions of the "dirty boson model", experiments show an
important variation in the values of the critical sheet resistance at the
transition $R_c$
\cite{Bielejec2002,Markovic1999,Steiner2004,Baturina2005,Okuma1998,Yazdani1995}.
Within one system, $R_c$ can vary between 2000 $\Omega$ to 9000
$\Omega$ \cite{Bielejec2002} depending on the applied magnetic field
or the normal resistance of the sample. Theories introducing a
fermionic channel of electronic conduction have been developed to
explain the non-universality of $R_c$ \cite{Yazdani1995} but these
are not entirely satisfactory since they do not account for values
of $R_c$ larger than $R_Q$ \cite{Markovic1999}.

    As one can see, all the experimental realizations of the SITs in thin disordered
films show a large variation in the measured critical exponents, as
well as in the critical resistance. This has led to the questioning
of the "dirty boson model". Some have suggested a percolation-based
mechanism \cite{Dubi2004}, others the contribution of fermions to
the conduction near the transition \cite{Yazdani1995}. Moreover, the
flat $R(T)$ curves found near the transition have put into question
Fisher's picture of an unique metallic separatrix between the
superconducting and insulating regimes. Some \cite{Das1999} have
suggested the existence of an intermediate metallic phase - the Bose
metal.

    In this context, it seemed to us particularly interesting to
provide another example of such transition. 2D Nb$_x$Si$_{1-x}$
films are interesting systems for this study. We have previously
shown that these films experience a magnetic field-tuned SIT
\cite{Aubin2006} with a product of critical exponents $\nu_H z =
0.67$, in agreement with other experimental data \cite{Markovic1998}
but in contradiction with the "dirty boson model". In this paper, we concentrate on the
thickness-driven SIT in this compound. The following sections will
be organized as follows : first, section \ref{sec:expmt} will detail the
experimental procedures. Section \ref{sec:dSIT} will explain the
finite-size scaling method we have used to analyze our results concerning
the disorder-induced transition
under non-zero magnetic fields and show that we have obtained
surprisingly small critical exponents for the transition. Combining
this analysis with our previously obtained results \cite{Aubin2006}, we infer the
phase diagram for Nb$_x$Si$_{1-x}$ (section \ref{sec:diag_phase}).
Finally, section \ref{sec:Discussion} will provide a
discussion on the interpretation of these sets of experiments on
disordered superconducting thin films and on the domain of validity
of the "dirty boson model".


\section{Experimental Procedure}
\label{sec:expmt}

The NbSi films have been prepared under ultrahigh vacuum by e-beam
co-deposition of Nb and Si. A series of four samples with
stoichiometry Nb$_{0.15}$Si$_{0.85}$ and thicknesses of 100, 50, 25
and 12.5 nm have been deposited onto sapphire substrates coated with a 50
nm-thick SiO underlayer. The films were synthesized during a single run in order to have
the samples' niobium concentrations as similar as possible. We
also took special care over the control of the sample's parameters :
the evaporation was controlled in situ by a special set of piezo-electric quartz in order to precisely monitor
the composition and the thickness of the deposition. These
two characteristics were then controlled ex situ by Rutherford Back
Scattering (RBS) and the results well fitted with the in situ
monitoring. Samples of the same stoichiometry with thicknesses down to
2.5 nm have been characterized by Atomic Force Microscopy
and showed no sign of morphological granularity nor inhomogeneity. The
superconducting transitions of these samples in zero magnetic field
are a few tens of mK sharp and show no sign of reentrant behavior as
usually observed for granular systems. Besides, all samples showed
the same resistivity at high temperature within 4\%. All these arguments lead
us to think that our samples are homogeneous in composition, non granular and only
differ from one another by their thickness. This conclusion is corroborated by a TEM
study \cite{Querlioz2005} showing that only Nb$_{x}$Si$_{1-x}$ alloys annealed at
500°C present Nb-rich clusters. The electrical
characteristics of the four films were measured down to 150 mK
using a dilution refrigerator. A perpendicular magnetic field could be
applied and was made to vary between 5 and 11 kOe. Resistance measurements were
performed using a standard AC lock-in detection technique operated at 23 Hz. A current
of 100 nA was applied to the sample, which is within the linear regime of
the $I$-$V$ characteristics for the considered films. All electrical leads
were filtered from RF at room temperature.


\section{$d$-induced transition}
\label{sec:dSIT}

    Before describing the renormalization procedure we have used
and the results thus obtained,
let us establish the dimensionality of our samples. In our
system, the mean free path $l$ is of the order of the interactomic
distance : $l \simeq$ 2.65 {\AA} \cite{Hucknall1992} and hence much
smaller than the superconducting coherence length $\xi_0$ given by
the BCS theory ($\xi_{0} = 0.18 \frac{\hbar v_F}{k_B T_{c0}}$ where
$v_F$ is the Fermi velocity estimated to be 2$\times$10$^8$ cm.s$^{-1}$
\cite{Marnieros1998}). In the "dirty" limit the effective coherence
length of the system is given by $\xi_{eff}=\sqrt{\xi_0
l}$. We also have to consider the dephasing length which acts near
the SIT as a cut-off length due to the finite temperature
\cite{Gantmakher2000ter,Fisher1990,Fisher1990bis} : $L_{\Phi} = \frac{\hbar^2}{m_e k_B
\xi_{eff}T}$ where $m_e$ is the mass of the electron. The smallest
length between $L_\Phi$ and $\xi_{eff}$ hence determines the
dimensionality of the film. The different relevant lengths are given
in table \ref{tab:gdeur_TSI}. The films with thicknesses ranging
from 12.5 to 50 nm can be considered to be 2D, whereas the 100 nm
film is 3D. In the renormalization procedure we shall focus on the
2D films, so that the resistances mentioned below are
sheet resistances. Let us also note that, in what follows, we used
the usual convention found in the SIT-related literature \cite{Markovic2001}:
the term "superconducting"
applies to curves that have a positive Temperature Coefficient of
Resistance (TCR : $\frac{dR}{dT}$), and, by contrast, we shall label as
"insulating" all curves having a negative TCR.

\begin{table}
\caption{\label{tab:gdeur_TSI} Relevant parameters for our samples :
the thickness $d$, the superconducting transition temperature $T_{c0}$, the BCS coherence length $\xi_0$,
the effective coherence length $\xi_{eff}$ and the dephasing length $L_\Phi$ at 0.3 K.}
\begin{ruledtabular}
\begin{tabular}{ccccc}
$d$ [nm] & $T_{c0}$ [mK] & $\xi_0$ [$\mu$m] & $\xi_{eff}$ [nm] & $L_\Phi$(0.3 K) [nm]\\
\hline
12.5 & 213 & 12.8 & 58.2 & 50\\
25 & 347 & 7.9 & 45.7 & 64\\
50 & 480 & 5.7 & 38.9 & 75\\
100 & 530 & 5.2 & 37.1 & 79\\
\end{tabular}
\end{ruledtabular}
\end{table}

    As shown by the rarity of experimental data concerning the
thickness-induced SIT, it is difficult to obtain a series of samples that are identical except for their thickness : unlike the magnetic
field, $d$ cannot be tuned continuously. We have therefore developed an analysis method which enables us to interpolate the system's
transport behavior between the discrete values of $d$ we experimentally have access to.

\begin{figure}
\graphicspath{{Programmes_figure/}}
\includegraphics [width=0.4\textwidth]{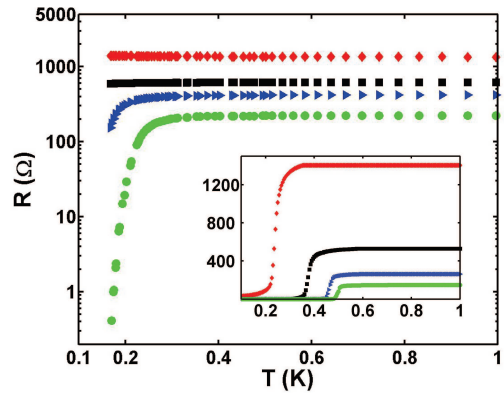}
\caption{\label{fig:RT} Resistance per square as function of temperature for
$H$ = 6.8 kOe. The curves for all four samples are represented. For this
particular value of the magnetic field, the 25 nm,
50 nm and 100 nm-thick films are superconducting, whereas the 12.5 nm-thick film
is insulating. Inset : The same data at zero magnetic field.}
\end{figure}

\begin{figure}
\graphicspath{{Programmes_figure/}}
\includegraphics [width=0.4\textwidth]{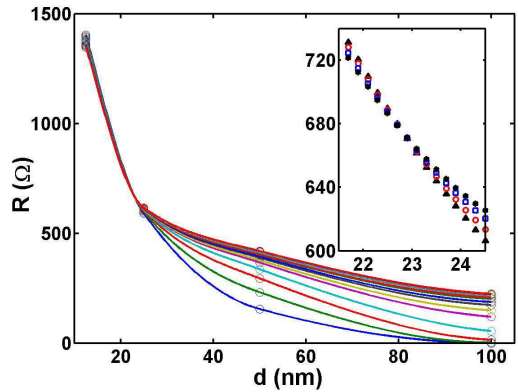}
\caption{\label{fig:Rd} Resistance per square as function of sample thickness for
H = 6.8 kOe. The curves are represented for 16 different values of the temperature
between 168 and 831 mK.
Insert : the same data are shown around the crossing point at about $d_c$ = 23 nm for
four particular temperatures : $T$ = 168, 186, 239 and 831 mK. This crossing point
is interpreted as the signature of a QPT.}
\end{figure}

    All four samples were superconducting at zero magnetic field (insert of figure \ref{fig:RT}) and were
progressively tuned through the transition by a finite $H$.
For each value of $H$, all four samples were studied (figure \ref{fig:RT})
and the diagram ($R$,$d$) traced for different temperatures presents a crossing
point (figure \ref{fig:Rd}). This is the signature of the QPT \cite{Markovic1999} and
allows us to estimate the critical thickness $d_c$ associated to the
magnetic field $H$. We repeat this process for all
values of $H$, obtaining a collection of critical
parameters couples ($d_c$,$H$).

    When $H$ is fixed, the thickness-induced transition is solely
governed by the distance to the transition $\delta_d =
\frac{|d-d_c|}{d_c}$. If these $d$-driven
transitions all belong to the same universality class, independent of the
particular value of $H$, the only relevant parameter for the
scaling of all our data is the value of $\delta_d =
\frac{|d-d_c(H)|}{d_c(H)}$. This means that \emph{all} curves
$R(\frac{|d-d_c(H)|}{d_c(H)},T)$ should collapse on two
universal curves. Note that the renormalized quantity we consider is $R$
and not $\frac{R}{R_c}$ as in \cite{Markovic1999} for we do not
find an universal critical resistance \cite{Cha1991}. For each individual sample,
this means that by tuning $H$, $d_c$ is made to vary and so does $\delta_d$.
In other words, the thickness $d$ being fixed, the critical thickness $d_c$
is changed via the magnetic field. Since the only relevant parameter for the scaling
is the distance $\delta_d$ to the transition, this situation is
ultimately equivalent to having a fixed critical
thickness and variable sample thicknesses (as in \cite{Markovic1999} for
example).

\begin{figure}
\graphicspath{{Programmes_figure/}}
\includegraphics [width=0.4\textwidth]{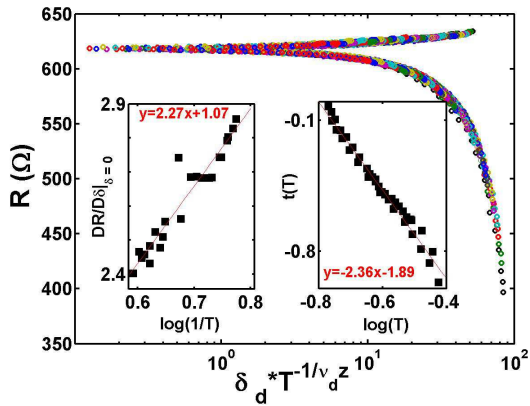}
\caption{\label{fig:renorm_250}  Renormalization of the resistance $R$ for
the critical exponents $\nu_d z = 0.4$ for the 25 nm-thick sample. Each
color is affected to a particular value of $\delta_d$. Left insert :
determination of the critical exponents product by the derivation method. Right
insert : determination of the critical exponents product by the $t(T)$ minimization
method.}
\end{figure}

    For each sample, the results were analyzed using two
independent scaling methods \cite{Markovic1999, Yazdani1995}.
First, for the derivative method, we plot $\frac{D R}{D \delta_d}|_{\delta_d=0}
\propto R_{c}T^{-\frac{1}{\nu_d z}}f'(0)$ as function of $\frac{1}{T}$
which, in a log-log diagram, gives a straight line of slope $\frac{1}{\nu_d z}$
(left insert figure \ref{fig:renorm_250}).
The second method consists in numerically finding $t(T)$ such that
$R(\delta_d,t(T)) = R_c f(\delta_d t(T))$ and that $t(T)$
yields the best collapse between the data measured at the temperature $T$
and the data measured at our lowest temperature (150 mK). To obey the scaling
law (equation \ref{eq:loi_échelle_B_d}), $t(T)$ should be of the form
$T^{-\frac{1}{\nu_d z}}$ and we can hence infer the value of $\nu_d z$
(right insert figure \ref{fig:renorm_250}).

    For all 2D samples, we obtained a product of critical
exponents of \boldmath$\nu_d z = 0.4 \pm 0.15$\unboldmath \; . We can check
this value of the exponents product by plotting $R$ as function of
$\delta_d \ast T^{-\frac{1}{\nu_d z}}$ (figure \ref{fig:renorm_250})
for the 25 nm-thick sample.
All data superimpose nicely in the ranges $0.16\leq T \leq
0.35$ K and $\mid \delta_d \mid \leq 1$, forming two curves only,
one representing the superconducting behavior and the other the
insulating side of the transition. $\mid\delta_d\mid$ = 1 still exhibits a critical
behavior since the corresponding data collapse on the same curves.
It is quite surprising that the scaling continues to
work that far from the critical point. The analysis performed on the 12.5 nm and the
50 nm-thick samples gave the same value of the product $\nu_d z$
within the uncertainty.

    This far, we have only considered the renormalization of
the resistance for one
sample at a time. In order to compare the critical behavior of the
different samples, we have to take into account their different
normal resistances. We therefore have to compare the quantity
$\frac{R}{R_n}$ where $R_n$ is the resistance taken at
high temperature, typically at 1K. This procedure is not usual in the literature and directly derives from the fact that,
in our experiment, $R_c$ is not universal and varies over one order of magnitude (see section V). The scaling of
$\frac{R}{R_c}$ then has no significance \cite{Cha1991}.

\begin{figure}
\graphicspath{{Programmes_figure/}}
\includegraphics [width=0.4\textwidth]{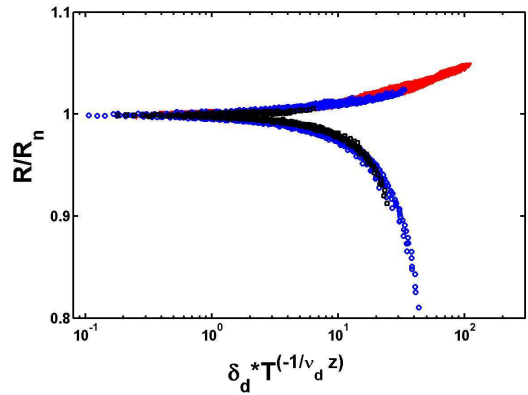}
\caption{\label{fig:renorm_all}  Renormalization of the renormalized resistance
$\frac{R}{R_n}$ for the critical exponents $\nu_d z = 0.4$ for the 12.5 (triangles),
25 (circles) and 50 (squares) nm-thick samples.}
\end{figure}

    We then looked for a critical exponent product that allowed
\emph{all} curves from \emph{all} samples to collapse. For each sample, we
adjusted the non-universal parameter $\alpha$ of equation (\ref{eq:loi_échelle_B_d})
for the curves to superimpose. We found $\alpha_{12.5nm}$ = 1.9,
$\alpha_{25nm}$ = 0.9, and $\alpha_{50nm}$ = 0.5 for a product of
\boldmath$\nu_d z = 0.4 \pm 0.1$\unboldmath. The corresponding criteria for the
renormalization are then very clearly defined : i. the magnetic
field was made to vary between 5.1 and 10.5 kOe by increments
of 0.1 kOe ; all critical points ($d_c$,$H_c$) corresponding to these fields have been
taken into account ; ii. the only constraint on the distance to the transition
is $\delta < 0.8$ ; iii. 0.17 $< T < $0.39 K. The result of the renormalization
is given figure \ref{fig:renorm_all}. This graph is particularly remarkable : even if our
samples have normal resistances varying by nearly one order of magnitude,
the corresponding resistances all collapse on a single renormalization
plot.

\section{Phase diagram}
\label{sec:diag_phase}

\begin{figure}
\graphicspath{{Programmes_figure/}}
\includegraphics [width=0.4\textwidth]{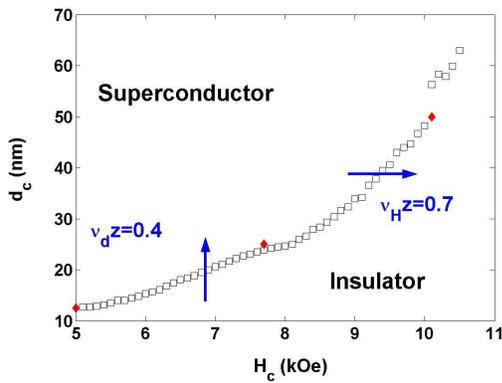}
\caption{\label{fig:diag_phase} Phase diagram for a-Nb$_{15}$Si$_{85}$ in
the ($H_c$, $d_c$) plane. The open symbols were obtained from the thickness-tuned
SIT, whereas the full symbols were obtained in \cite{Aubin2006} for the magnetic-field
tuned transition.}
\end{figure}

    The renormalization method has enabled us to measure a number of critical
parameters couples ($H_c$,$d_c$) although we only had four different samples. We can
hence draw part of the phase diagram for Nb$_x$Si$_{1-x}$ thin films (figure
\ref{fig:diag_phase}). The line formed by the critical points
separates an insulating region at high fields
and small thicknesses form a superconducting region at low field and large
thicknesses. Of course these critical points coincide with those determined from
the magnetic field-induced SIT \cite{Aubin2006}. As for a-Bi \cite{Markovic1999},
depending on the parameter tuned to cross this line, the critical exponents product
found is different : $\nu_H z$ = 0.7 when the field is varied,
whereas a variation of the sample's thickness gives $\nu_d z$ = 0.4. We thus
confirm that these two SITs belong to two separate universality classes.

\section{Discussion}
\label{sec:Discussion}

    First let us comment on the value found for the critical exponents
product. For a-NbSi in a thickness-induced SIT, we have found $\nu_d z = 0.4$.
This value is surprising when compared to other critical exponents found by
other groups, thin a-Bi films for instance, for which $\nu_d z = 1.4$. At this
point we do not have any clear explanation for this important difference.
However $\nu_d z = 1.4$ is close to what is predicted for classical 2D
percolative systems ($\nu_d = 4/3$) and a-Bi thin films present a thickness-induced
SIT for very shallow thicknesses (a few angstr\"{o}ms, 20 {\AA} at most). In this
sense also, our system is particularly interesting since it allows 2D samples
to experience a thickness-driven SIT at reasonable thicknesses where the
roughness of the film, the surface state of the substrate or the microscopic
details of the film's growth should not be important factors. $\nu_d z =0.4$ is
also surprisingly small considering the theoretical predictions that have been
made to this day \cite{Sondhi1997}. Although the exact value of this product might be affected
by the uncertainty on the determination of the exponents ($\pm 0.1$) and by
the small number of samples we have, at any rate, we can confidently say that
$\nu_d z < 1$ which is inconsistent with the "dirty boson model". If we assume that
$z=1$, the consequence of this is that $\nu_d < 1$. Many authors have
pointed to the fact that this violates the so-called "Harris criterion" ($\nu>2/d$) \cite{Chayes1986}, however
this criterion is valid for small disorder and since our system consists in
amorphous films in which the mean free path is of the order of the inter-atomic distance,
it is not all that shocking that the value found for the localization length exponent
does not obey this inequality \cite{Avishai_perso}.

\begin{figure}
\graphicspath{{Programmes_figure/}}
\includegraphics [width=0.4\textwidth]{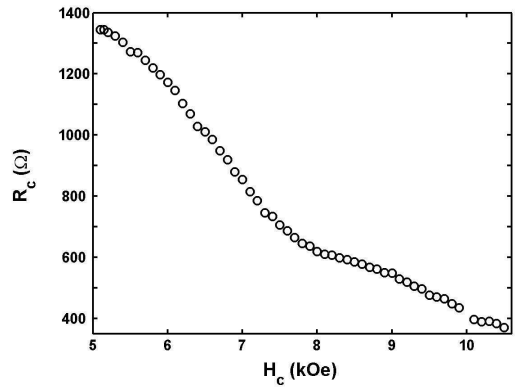}
\caption{\label{fig:Rc} Critical resistance as function of the critical field
for a-NbSi films.}
\end{figure}

    Another point that has much been discussed related to the "dirty boson model"
is the value of the critical sheet resistance. In this set of experiments, we show that $R_c$ varies over a large
range when either the magnetic field or the thickness is varied (figure \ref{fig:Rc}).

\begin{figure}
\graphicspath{{Programmes_figure/}}
\includegraphics [width=0.4\textwidth]{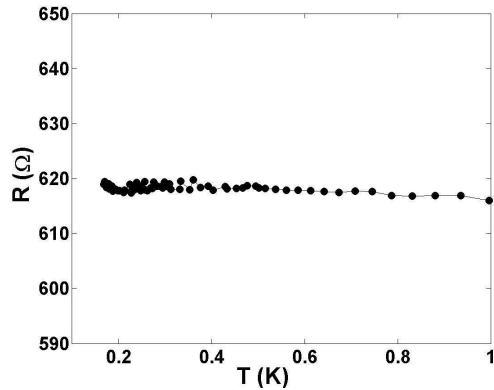}
\caption{\label{fig:sat} Resistance as function of the temperature for the 50 nm-thick
sample at $H$ = 7.9 kOe. Over one decade variation in temperature, the film's resistance
only varies within 3.5 $\Omega$ (0.5\% in relative value), which is our experimental
uncertainty in this range.}
\end{figure}

    Until now we have analyzed our results by comparing them to the
"dirty boson model". Although the renormalization procedure works
remarkably well for all systems studied to date - which means the SIT
is indeed a QPT \cite{Sondhi1997} -, two important predictions of this model
($\nu > 2/d$ and $R_c=h/4e^2$) are not verified by a-NbSi thin films as well as
in other systems (a-Bi \cite{Markovic1999}, a-Be \cite{Bielejec2002}, NdCeCuO
\cite{Gantmakher2003}, MoGe \cite{Mason2002}, InOx
\cite{Hebard1990,Steiner2004}, TiN \cite{Baturina2005},
MoSi \cite{Okuma1998}...). One might therefore put this
model into question. Tunneling effect experiments suggest \cite{Dynes1986,Valles1989,Valles1992}
that, for homogeneous systems, amplitude fluctuations of the order parameter play a role even in
the vicinity of the SIT : when the films' thickness decreases, the superconducting
gap $\Delta$ and the critical temperature decrease together, monotically, such that
$2\Delta / T_c \simeq constant$. In this picture, near the SIT, the
amplitude of the superconducting order parameter can become very small, whereas
an essential point in the "dirty boson model" is that its amplitude is finite
near the transition. The same studies show that, even in the "superconducting"
- in the previously-defined sense of the TCR - region, the one particle
density of state is not zero, meaning that there
are normal excitations coming from electrons that are not involved in any
Cooper pair. This would mean that amplitude fluctuations of the system must
be taken into account for a correct description of the transition, which is not
the case in M.P.A. Fisher's model. The suggestion by some authors that other phase(s) may be involved
in  between the superconducting and the insulating regimes is particularly
interesting. Some have suggested a vortex-liquid phase \cite{Chervenak1996,Okuma2005} which has recently
\cite{Anderson2007} been linked to the problem of anomalous Nernst effect in the cuprates.
As recent measurements on amorphous superconductors have
shown \cite{Pourret2006,Pourret2007,Spathis2008}, Nernst effect is a very sensitive probe
of amplitude fluctuations \cite{Pourret2006,Pourret2007} and phase
fluctuations \cite{Spathis2008} of superconducting order parameter. These last works
suggest that measurements of the Nernst effect should be a relevant probe to
test the existence of this vortex liquid phase.
However, there has not been clear predictions on how
the thickness variation should affect this phase. Also very appealing is the
suggestion that there is a bosonic metallic phase, such as the Bose metal
\cite{Das1999}, involved. This hypothesis is very interesting, in particular
when one takes a close look at the resistive
behavior of our films. Indeed, at low temperatures, the resistance
of some samples seem to saturate at a finite value (figure \ref{fig:sat}),
displaying a large temperature range where the resistance is independent
of the temperature. However a study at lower temperatures should
be undertaken to confirm this tendency. Let us restate that the qualification of
insulating or superconducting have been arbitrarily attributed to
$\frac{\partial R}{\partial T} < 0$ (resp. $\frac{\partial R}{\partial T} > 0$)
curves without any other ground than the assumption made by the "dirty boson
model" that only these two phases existed. All these arguments (the amplitude
fluctuations of the order parameter, a possible fermionic channel, the
suggestion of a Bose metal...) plead in favor of a reconsideration of the
"dirty boson model" and further experimental investigations of these systems.

    In conclusion, we have studied the thickness-induced SIT in the presence of a
perpendicular magnetic field on a-Nb$_{15}$Si$_{85}$ thin films of
thicknesses ranging from 12.5 to 100 nm. We have found the signature of a QPT
when the sample thickness is lowered. The corresponding critical exponents
product is $\nu_d z \simeq 0.4 \pm 0.1$. This value is different from the one
found in the analysis of the magnetic field-induced transition in the same
compound for which $\nu_H z = 0.65$. These two SITs therefore belong to two
different universality classes. However the very small value of $\nu_d z$ cannot
be explained by the existing models for this transition. Further experimental
investigations are needed to understand the growing discrepancies between the various
experimental results and between these results and the theory.



\begin{acknowledgments}
We are grateful to acknowledge stimulating discussions with M.V. Feigelman, S. Okuma, C. Chapelier,
and Y. Avishai.
\end{acknowledgments}


\newpage 


\bibliography{TSI_d_B_brouillon_v9}


\end{document}